\documentclass[twocolumn,pra,letterpaper,showpacs,superscriptaddress,floatfix]{revtex4}
\usepackage{graphicx,psfrag,bbm,latexsym,color,dcolumn,bm,dsfont,bbm,color,mathrsfs,bbold,latexsym,amsmath,amsfonts,amssymb}

\newcommand{\beq}{\begin{equation}}
\newcommand{\eeq}{\end{equation}}

\newcommand{\beqa}{\begin{eqnarray}}
\newcommand{\eeqa}{\end{eqnarray}}

\newcommand{\bw}{\begin{widetext}}
\newcommand{\ew}{\end{widetext}}

\newcommand{\CZ}{\mathcal Z}

\begin{document}
              
\title{Spontaneous Parity Violation}
\author{P.R. Crompton}
\affiliation{
Institut f\"ur Theoretische Physik, Universit\"at Leipzig, \\
Augustusplatz 10/11, D-04109 Leipzig, Germany.}
\vspace{0.2in}
\date{\today}
\email{crompton@itp.uni-leipzig.de}

\begin{abstract}
\vspace{0.2in}
{We disprove the Vafa-Witten theorem on the impossibility of spontaneously breaking parity in vector-like 
gauge field theories, identifying a mechanism driven by quantum fluctuations. With the introduction of a 
meromorphic Lattice formulation, defined over 5 dimensions, we demonstrate that the minima of the free 
energy can be distinct from the maxima of the partition function : identifying and 
evaluating a suitable contour for the partition function defined such that asymptotic behaviour of the 
complex action is non-oscillatory.}
\vspace{0.1in}
\end{abstract}


\maketitle

Recently attempts have been made to determine the properties of the Grand Canonical 
Partition Function of QCD at nonzero baryon densities using numerical Lattice QCD 
approaches \cite{revmu}. 
These results are directly relevant for determining the equation of state of QCD for experimental Heavy Ion Collider 
results. These efforts are frustrated  numerically by the difficulties in constructing 
efficient sampling algorithms at finite baryon densities. One faces the problem of a 
complex-valued Lattice action, whose phase oscillates rapidly as a function of the configuration within a Lattice 
statistical ensemble. This then prevents the effective use of importance sampling methods. All recent 
sampling improvement treatments in this vein appear now to suffer from the same generic flaw of an 
unphysical light pseudo-baryonic transition \cite{kim,immu}. 

To gain understanding of this feature an 
analytic treatment of partition function zeroes for QCD and QCD-like theories with a complex 
action problem was recently proposed \cite{vafaa,vafab}. Here the properties of 
the zeroes are identified as relating directly to specific regions of parameter space 
in the continuum where the partition function is itself ill-defined. This is linked in the 
discussion to the Vafa-Witten Theorem on the impossibility of spontaneously breaking parity 
symmetry in QCD \cite{VW}. A local bosonic parity breaking source term in the action is shown to generate a similarly 
pathologically ill-defined continuum partition function. This is demonstrated explicitly through 
the properties of the zeroes, and is used to justify the main theorem result. 
Although not included in this specific zeroes analysis the Vafa-Witten theorem result is 
understood to represent a wider pathology in Lattice simulation \cite{cohen,blaze}. Bosonic terms with substructure 
in the Vafa-Witten formalism give rise to unphysical diagrammatic contributions to the Lattice QCD 
partition function. For example, in $N_{f}=2$ QCD with a four fermion operator source \cite{sharpe}, 
or the Aoki phase with Wilson fermions \cite{aokia,aokib,aokic}.

We now apply a new quantum system partition function zeroes treatment to this problem, and show 
that in contrast {\emph {it is}} possible to give a well-defined continuum partition function 
for {\emph {all}} generic complex action problems (of those of the form in the Vafa-Witten discussion) for 
{\emph{all}} parameter space regions. Under the proviso, that, the local source term has a highly nonlocal 
component for which the continuum limit has been reached over an additional phase space dimension. We introduce 
a new quantum loop operator definition for the partition function and show that it is formally equivalent to the 
Vafa-Witten construction, but in addition, meromorphic. 
This new analysis is used to rigorously disprove the Vafa-Witten result, and demonstrate that it is in fact 
possible for any global Lattice symmetry (like taste) to be spontaneously broken via the system dynamics. 
Specifically, we demonstrate that the branch cut argument introduced in \cite{vafaa,vafab,aguado,ji} 
used to argue that the partition function is ill-defined is specious. The branch in fact can be resolved 
analytically and unambiguously for all Lattice system sizes, $N$, and in particular in the asymptotic limit. 

The new analysis also indicates that unless the Lattice simulation of a quantum system with a 
complex action problem produces extremely finely balanced statistical averaging between degenerate vacua 
(namely that which in the asymptotic limit $N \rightarrow \infty$ connects smoothly with the continuum) then spurious 
phase behaviour will be observed. We have applied a new fine-balancing scheme recently, based on 
this understanding, via numerical 
simulation to several Lattice QCD models at finite baryon densities and independently quantum spin 
systems (which have a hidden continuous gauge symmetry) \cite{su2a,su2b,immu,Deltaf,n=4}. 
This article then, in part, provides a generic 
proof of the existence of the continuum limit of these ongoing Lattice analyses, and the validity 
of our generic continuum matching program for these calculations.

The purpose of this new article is to establish the origin of large 
light disconnected contribution in recent lattice QCD schemes at finite baryon densities 
in a generic manner. Specifically, we wish to establish the relationship between the statistical ambiguities 
of numerical sampling and the parity symmetry violation arguments of Vafa and Witten. This is 
expressed through a saddle point point discussion of the partition function and is specific to 
our new quantum loop operator formulation.
To do this, following Lee and Yang \cite{lee+yang}, we wish to now argue for separate treatment for the approach 
to the continuum limit of the zeroes polynomial in quantum systems, as opposed to classical 
systems. We therefore rederive the partition function zeroes polynomial proof of Lee and Yang and 
the validity of its approach to the continuum limit for the quantum case, formulated in terms of 
loop operators. In doing so, we demonstrate that 
this is the generic origin of these disconnected contributions in numerical simulations. This then 
has direct applicability in the design of new improved algorithms - to include correct quantum 
effects - both for Lattice QCD and quantum spin systems that are otherwise spoilt by multiplicative 
nonrenormalizable anomalies induced through the noncommutativity of finite Lattice system dynamics \cite{renorm,asymp}. 

This article is then organised as follows :
in Section \ref{sec1} we introduce a new quantum loop operator formalism for the partition function 
and identify that this is formally equivalent to the local bosonic operator expansion on finite 
system size introduced in the Vafa-Witten discussion. We then consider the continuum limit of 
this formulation and demonstrate that the zeroes of the loop operator polynomial are formally 
equivalent to partition function zeroes, but only as defined with reference to a prior noncommuting 
continuum limit being taken over an additional dimension in an extended phase space. 
We then derive 
the saddle point solution for the partition function of this complex-valued quantum loop operator
formulation.

In Section \ref{sec2}, we discuss the analytic continuation of the saddle point solution with reference 
to the symmetry breaking properties of the newly defined quantum partition function, and the Vafa-Witten 
discussion. We demonstrate that the results can be analytically continued across the entire phase plane in the 
zeroes expansion parameter. Therefore any nonperturbative operator satisfying the quantum loop operator definition can 
be analytically continued throughout the complex phase-plane, and is an exact analytic solution to the complex
action problem.

\section{Quantum Loop Operators}
\label{sec1}
The partition function, $\CZ$, for a general continuous group $G$ we define in terms of quantum loop operators by,

\beq
\label{partitionz}
\CZ_{N}(t) = \int {\rm{d}}g \,\,\,  
{\rm{exp}}\!\left[\,\int_{0}^{t} A_g(\bm{n}_s)\,z-V(\bm{n}_s) \,\,ds \,\right],
\eeq
with $t\in \mathbb{R}, \,\,\, z \in \mathbb{C}$,

\beqa\label{Hilberta}
A_g(\bm{n}) & \equiv & 
\sum_{(i,j)}^{N \otimes T} \,\, \sum_{\sigma \in G} 
\lambda_{ij\sigma}(\bm{n})\frac
{\langle \bm{n}\oplus \bm{1}_{i\sigma} \oplus \bm{1}_{j\sigma}| g \rangle}
{\langle \bm{n}| g \rangle}\,, \\
\label{Hilbertb}
V(\bm{n}) & \equiv & 
\sum_{(i,j)}^{N \otimes T} \,\, \sum_{\sigma \in G} 
\lambda'_{ij\sigma}(\bm{n})\frac
{\langle \bm{n}\oplus \bm{1}_{i\sigma} \oplus \bm{1}_{\sigma j}| \bm{n} \rangle}
{\langle \bm{n}| \bm{n} \rangle}\,,
\eeqa
where $\lambda_{ij\sigma}(\bm{n}),\lambda'_{ij\sigma}(\bm{n})\in \mathbb{C}$, $\{\sigma\}$ form some finite subset of the 
elements of $G$, and $g$ is a general element of $G$. We treat the complex-valued continuous symmetry generalisation of the problem considered in 
\cite{presilla-gauge,presilla-old}.
The fact that the operator definition is diagonalisable implicitly defines a local gauge transformation on the Hilbert 
space, 

\beqa
\label{gauge} 
\sum_{(i,j)}^{N \otimes T}\!\!\!\!\!\! & & \!\!\! \sum_{\sigma=G} 
\lambda_{ij\sigma}(\bm{n})\frac {\langle \bm{n}'\oplus \bm{1}_{i\sigma} \oplus \bm{1}_{j\sigma}| \bm{n} \rangle}
{\langle \bm{n}| g \rangle} \langle \bm{n}' | g \rangle \nonumber \\
& = & \langle \bm{n}'|g\rangle \, \langle\bm{n}'| 
A |\bm{n}\rangle \, {\langle \bm{n}|g\rangle}^{-1} = \langle \bm{n}'|A_g  |\bm{n}\rangle\,. 
\eeqa
We should compare this directly with the single plaquette model given in terms of the Vandermonde determinant 
discussed in \cite{vandermondea,vandermondeb}, which yields a similar form of time-ordered product. The connection to Vafa-Witten formalism is made 
explicit by noticing that equivalently the integration of the exponent in can be performed over the spatial Lattice 
volume via a change of basis, $\bm{n}_s'$. The partition function is then given in the usual Vafa-Witten form,   

\beqa
\label{partitiont}
\CZ_{T}(z) & = & \int {\rm{d}}g \,\,\, 
{\rm{exp}}\!\left[\,\int_{0}^{z} A_{g}(\bm{n}_{s'})\,t-V(\bm{n}_{s'}) \,ds'\,\right]\nonumber \\
& = & \int {\rm{d}}g \,\,\, 
{\rm{exp}}\!\left[\,\int_{0}^{z} A'_{g}(\bm{n}_{s'})\,it-V'(\bm{n}_{s'}) \,ds'\,\right]
\eeqa
\beqa
A'_{g}(\bm{n}_{s'}) & = & {\rm{Im}}[A_{g}(\bm{n}_{s'})] \,,\\
V'(\bm{n}_{s'}) & = & V(\bm{n}_{s'}) - {\rm{Re}}[A_{g}(\bm{n}_{s'})]\,,
\eeqa
where $V(\bm{n}_{s'}) \in \mathbb{R}$ since it is hermitean in the previous basis, $\bm{n}_s$. 

\subsection{Thermodynamic Limit Monotonicity}

It is surprisingly simple to prove the existence of the thermodynamic limit for partition function in (\ref{partitionz}). 
The proof consists of two parts. The first step is to identify the saddle point solution of the partition function. 
The second step is then to determine the limiting value of this asymptotic solution 
when the neighbourhood of the saddle point is extended to cover the entire phase space. Apart from isolated 
singularities the partition function is then necessarily holomorphic. This is the essence of Lee and Yang's 
discussion in \cite{lee+yang,poissona}. 

For the quantum case we face now an additional subtlety. This has not been previously addressed in the literature.
Partition functions for quantum systems are defined over an extended phase space - a trace is taken over an 
additional dimension. Two examples are given above in (\ref{partitionz}) and (\ref{partitiont}). 
For a quantum system to be truly holomorphic, and the 
thermodynamic limit exist, it is therefore not merely sufficient that the thermodynamic limit be proven to exist 
over either $\bm{n}_{s}$ or $\bm{n}_{s'}$. Rather, the thermodynamic limit must exist over $\bm{n}$. 
Expressing this another way, existing 
partition function studies invoking this holomorphic property over $\bm{n}_{s'}$ (holomorphic in $z$) have a 
thermodynamic limit that is strictly speaking only defined up to infrared divergences (ie. they are not necessarily simultaneously 
holomorhpic in $t$). 
For Lee and Yang with classical systems, which have no trace relation, this complication does not arise.
The partition function in this case is given directly over $\bm{n}$, without the time-ordering of the integrand.

We now prove the existence of the thermodynamic limit for the new quantum loop operator formulation for the partition 
function in (\ref{partitionz}). 
The saddle point, which maximises the partition function in some region of the extended phase space $M$, is defined through,
\beq
R_{M} \equiv \sup_{M} R(\bm{m})\,, 
\eeq
where $\bm{m}$ is parameterises some contour in $N \otimes T$, and $R$ is the integrand of the partition function. The 
value of the integrand in the neighbourhood of the saddle point, $M_{\delta}$, is defined through, 
\beq
R_{\delta}(\bm{m}) \equiv R(\bm{m}) - R_{M} \leq 0 \,. 
\eeq
Not every point 
contained in $M$ is necessarily in the neighbourhood of the saddle point. We consider how the neighbourhood evolves 
as the contour is increased in length : the limit where $\bm{m} \rightarrow \bm{n}$. 

\beq
\label{subsupport}
\lim_{\bm{m} \rightarrow \bm{n}} R_{\delta}(\bm{m}) = \lim_{\bm{m} \rightarrow \bm{n}} R(\bm{m}) - R_{M} \leq 0\,.
\eeq
The relation between the consecutive basis elements (along $\bm{m}$) is given partly through the local gauge symmetry 
transformation in (\ref{gauge}). However, we can have a situation for some general group $G$ in which we have multiple degenerate solutions
of (\ref{subsupport}). 
In this case a subspace must be projected out to time-order the integrand and resolve this degeneracy. With this resolution 
is then necessarily monotonic and decreasing, since the potential nonlocal ambiguities are removed. 
Next, taking the limit $\lim_{M\to N \otimes T}$ we have, 

\beqa
\lim_{M\to N \otimes T} \!\!\!\! & & \!\!\!\! [ \,\, \lim_{\bm{m} \rightarrow \bm{n}} R_{\delta}(\bm{m}) ] \nonumber\\ 
& = & \lim_{M\to N \otimes T} \,\, [ \,\, \lim_{\bm{m} \rightarrow \bm{n}} R(\bm{m}) - R_{M} ] \leq 0\,.
\eeqa
This limit is also monotonic and decreasing with above choices, and therefore the thermodynamic limit of the quantum loop 
operator partition function in is well-defined. The holomorphic properties we are therefore interested in are those of the 
space $N \otimes T$, modulo, the time-ordering necessary to resolve the nonlocal ambiguities of the gauge group. It should be 
clear now from a numerical perspective that it is not necessary to perform either integral in the integrands exactly : in order 
to be simultaneously holomorphic in $t$ and $z$ it is sufficient for the integrand to cover $N \otimes T$.

\subsection{Saddle Point Determination}
We are interested to find an explicit expression for the evolution of the partition function, $\CZ_{N}$,
in the limit $N \rightarrow \infty, N > 0$. There are exactly $N$ singularities in the partition function in 
as a consequence of the above monotonicity arguments. By construction, the loop operator definition of $\CZ_{N}$ is 
meromorphic so we can equally well formulate the discussion in terms of the poles of $\CZ_{N}$, $\{V_{k}\}$, or 
the zeroes, $\{\lambda_{k}\}$. The poles turn out to be easier to treat than the zeroes, but for clarity the relation 
between the poles and zeroes is given through, 
\beqa
A(z) & \equiv & \prod_{k=0}^{N}\frac{z+\lambda_{k}}{z+V_{k}} \,, \\
\label{argument}
\frac{1}{2\pi i} \int_{\mathcal{C}} \frac{A'(z)}{A(z)} dz & = & 
\sum_{k} n({\mathcal{C}},\lambda_{k}) - \sum_{k} n'({\mathcal{C}},V_{k}),
\eeqa
where $n$ and $n'$ are the respective homotopies of the zeroes and poles, and ${\mathcal{C}}$ is some contour 
homologous to $0$ in $N \otimes T$. This property will become important shortly when we consider finite multiplicities for the 
singularities, or equivalently, global symmetries in $N \otimes T$.

The schema we now follow is the complex-valued generalisation of the problem considered in \cite{presilla-saddle,bessel}. Namely, we define the 
Laplace transform of the partition function, $\widetilde{\CZ}_{N}$, as a function of the poles, $\{V_{k}\}$. 
Then we formulate the inverse Laplace transform that gives us back the partition function expressed as a 
function of $t$, $N$ and $\{V_{k}\}$. Finally, we evaluate this integral 
equation for $\CZ_{N}$ via standard asymptotic series approaches in order to identify the dominant behaviour 
of the partition function in the above limit. In particular, this approach allows us to identify the saddle 
point solution of the partition function as distinct from the asymptotic convergence properties of the zeroes 
towards the ${\rm{Re}}\,z$ axis.

The Laplace transform of the partition function polynomial in at finite $N$ is given in terms of the poles 
of the partition function by, 
\beq
\widetilde{\CZ}_{N}(z) = 
\int_{0}^{\infty}dt e^{-zt}{\CZ}_{N}(t) = \epsilon^N\prod_{k=0}^{N}\frac{1}{z+V_{k}}, 
\eeq
where $z,\epsilon,V_{k} \in \mathbb{C}$.
The inverse Laplace transform of is given by, 

\beq
\CZ_{N}(t)=\frac{1}{2\pi i}\int_{\mathcal{C}} dz
e^{zt} \widetilde{\CZ}_{N}(z) =\frac{1}{2\pi i \epsilon}
\int_{\mathcal{C}} dz \, \exp\left[N \varphi(z) \right],
\eeq

\begin{eqnarray}
\varphi(z) =
\frac{zt}{N} - 
\sum_{k=0}^{N}\frac{1}{N} \log\left(\frac{z+V_{k}}{\epsilon}\right).
\end{eqnarray}
The contour ${\mathcal{C}}$ in, is defined as any suitable path running between $\pm {\rm{i}}\infty$, for which  
any singular behaviour is appropriately treated. 
We want to now evaluate the asymptotic behaviour of the partition function separately for the two cases : when 
a parity symmetry is, and is not, realised on the vacuum. 

In the first case we have no special symmetry relations defined between the poles, and $\varphi(z)$ is therefore 
of the generic form above in. Solving explicitly for $\varphi'(z_{0})=0$ we have,
\beq
\sum_{k=0}^{N} \, \frac{ 1 }{ z_{0} + V_{k} } = t\,. 
\eeq
Note, the location of the saddle point, $z_{0}$, is a function of both $\{V_{k}\}$ and $t$. Potentially the saddle point lies to 
the left of some subset of the poles, $\{V_{m}\}, \, m \leq N$. Therefore the integral in is given by, 
\vspace{-0.18in}
\beq 
\CZ_{N}(t)=\frac{1}{2\pi i \epsilon}
\int_{\mathcal{C}_{z_{0}}} \, \exp\left[N \varphi(z) \right] dz + \sum^{m}_{l=1} \frac{e^{V_{l}t}}{\epsilon} 
\prod_{k\neq l}^{N} \frac{1}{V_{l}+V_{k}}
\eeq
where $\mathcal{C}_{z_{0}}$ is the contour that passes through $z_{0}$. 
For this first case, when one assumes no parity symmetry is realised with the gauge fields, the asymptotic 
behaviour of the partition function is therefore given by, 
\bw
\beqa
\lim_{N\rightarrow \infty, N>0} \, \CZ_{N}(t) 
& = & \frac{1}{2\pi i \epsilon} \, \exp\left[ N \varphi(z_{0}) - \frac{i}{2}\theta(z_{0}) \right]
\left( \sqrt{\frac{-2\pi}{N|\varphi''(z_{0})|}} + O(N^{-3/2}) \right) , \\
& = & \exp\left[ z_{0}t - 
\sum_{k=0}^{N}\log\left(\frac{z_{0}+V_{k}}{\epsilon}\right) - \frac{i}{2}\theta(z_{0}) \right]
\left( \sqrt{\frac{1}{ 2 \pi \epsilon^{2} N |\varphi''(z_{0})| }}+ O(N^{-3/2}) \right),
\eeqa
\ew
where $\theta(z_{0}) = {\rm{Arg}} \, \varphi''(z_{0})$. 

For the second case it is straightforward to define a parity symmetry action on $Z_{N}(t)$ simply by reversing the orientation of the 
elementary plaquettes in $N\otimes T$. This then yields the symmetry of the zeroes $\{\lambda_{k},\lambda_{k}^{*}\}$. From the 
Argument Principle in (\ref{argument}) the poles are then given through,

\beq
\label{branch1}
\widetilde{\CZ}_{N}(z)  =  
\epsilon^N\prod_{k=0}^{N}\frac{1}{\sqrt{(z + V_{k})(z + V_{k}^{*})}} \,, 
\eeq

\beq
\label{saddle}
\varphi(z)  =  
\frac{zt}{N} - 
\sum_{k=0}^{N}\frac{1}{N} \log\left(\frac{\sqrt{(z + V_{k})(z + V_{k}^{*})}}{\epsilon}\right).
\eeq

The singularities at $z= {\rm{Re}}\,V_{k} \pm {\rm{Im}}\,V_{k}$ therefore correspond to branch points, 
with a branching exponent of $-1/2$. Again, we want to define the contour such that it passes through $z_{0}$, but 
this is again potentially obstructed by singularities lying to the right of $z_{0}$. We therefore choose the branch cut to 
correspond to this deformation over the saddle point. If ${\rm{Im}}\,V_{k} < 0$ the deformation will start at the branch 
point and descend to $-i\infty$. If the ${\rm{Im}}\,V_{k} > 0$ the deformation will start at the branch point and ascend to 
$i\infty$. If ${\rm{Im}}\,V_{k} = 0$ the singularity is a pole, which we will treat as a residue contribution as before.
This choice of branch cut does not introduce an arbitrariness into the calculation. Although it is not now possible to analytically 
continue across the cut, $\varphi(z)$ is specified unambiguously along the original contour, and the analytic continuation 
between contours is also specified unambiguously. Crucially we keep the same branch cut for the pairs $\{ V_{l}^{}, V_{l}^{*}\}$.
The cuts to the left and right of the contour deformation $\mathcal{C}$ are defined through,

\beq 
(z+V_{l})_{+}^{-1/2} = e^{-\pi i} (z+V_{l})_{-}^{-1/2}\,.
\eeq
The analogue of the residue contribution for the branch points is therefore,

\beqa
\CZ_{N}^{\,l}(t) & = & \int_{\mathcal{C}} e^{N \varphi(z, k\neq l)} (z+V_{l}^{*} )^{-1/2} (z+V_{l})^{-1/2}_{-} dz \nonumber \\
& - & \int_{\mathcal{C}} e^{N \varphi(z, k\neq l)} (z+V_{l}^{*} )^{-1/2} (z+V_{l})^{-1/2}_{+} dz \nonumber \\ \\
& = &  \,\,\, 2 \int_{\mathcal{C}} e^{N \varphi(z, k\neq l)} 
(z+V_{l}^{*} )^{-1/2} (z+V_{l})^{-1/2}_{-} dz \nonumber \\
\eeqa
\beq
\label{branch}
\sim 2 e^{N \varphi(z=V_{l}, \,k\neq l)} 
\sum_{n=0}^{\infty} \frac{\Gamma(n + 1/2\,)b^{n}(0)}{n! \,\, N^{\,n + 1/2\,}}
\eeq
where, 

\beq
b(0) = \left( -2i \, {\rm{Im}} \, V_{l}\, .\,\, \varphi'(V_{l}, k\neq l) \right)^{-1/2},
\eeq
\beq
\varphi'(z\!=\!V_{l}, k\neq l) = \!
\frac{t}{N} \!- \!\sum_{k=0, k\neq l}^{N}\frac{1}{N} \frac{ {\rm{Re}}\,(V_{l} + V_{k})  }{\sqrt{(V_{l} + V_{k})(V_{l} + V_{k}^{*})}} \, . 
\eeq
and the $b^{n}(0)$ form the analogue of the Laurent series for the branch points. 
The solutions are not oscillatory in $N$, as was found with previous discussions. In our new formulation we can see 
as well that the behaviour of the branch points can wholly dominate the asymptotic form of the partition function in the case in 
which a parity symmetry is realised, solely dependent on the relative magnitude of the prefactor, ${\rm{exp}}[\varphi(V_{k},t)]$.

We are able to reach this new point of unambiguous saddle point solutions for two reasons. Firstly, the new problem is defined over 
a higher dimension than previous discussions, so $z$ and $t$ are separated. This means that the singularities can be 
distinguished from asymptotic scaling. Secondly, because the new construction is meromorhpic we can identify all the 
singularities in $\CZ_{N}$, which are of a finite number. 

The above arguments also are readily generalised to cases with different 
branching exponents - corresponding to different global symmetries. Specifically, if the branching exponent is less than -1 then the 
above asymptotic identifications for the branches can be performed via integration by parts. 

\section{spontaneous symmetry breaking}
\label{sec2}

We have constructed in (\ref{partitionz}) a partition function that is formally equivalent to that of the Vafa-Witten discussion. It consists 
of a local bosonic path integral, on which is defined a symmetry breaking source term, $A'_{g}(\bm{n}_{s'})$. 
To relate our saddle point identification 
further to the Vafa-Witten discussion on spontaneous parity violation we now need to evaluate the vacuum expectation of the source 
term. Since the path integral is time-ordered this is given by, 
\beqa
\langle {\cal O} \rangle & = & \lim_{N\rightarrow \infty} \left( \frac{1}{N} \frac{\partial \log \left ( \CZ_{N}( t ) \right )}
{\partial t} \right)\\
& =  & \lim_{N\rightarrow \infty} \left( \frac{1}{N} \frac{ \int \!dz\,\,z
e^{zt} \, \widetilde{\CZ}_{N}(z)}{ \int \!dz \,\, e^{zt} \, \widetilde{\CZ}_{N}(z) } \right)\,.
\eeqa
We therefore have two possible solutions, from (\ref{saddle}) and (\ref{branch}), for the case when parity is realised by the system,

\beq
\langle {\cal O} \rangle = z_{0} \quad {\rm{or}} \quad \langle {\cal O} \rangle = V_{l}\,.
\eeq 
In particular we can see for this second case in (\ref{branch}), 
associated with the branch point, that the free energy density minima does not correspond
with the partition function maxima at $z_{0}$. What we can see through our new formalism is that at finite $N$ there is a finite 
subtraction to made to properly define the contour for the partition function, and this itself can give rise to a parity breaking contribution.
From (\ref{saddle}) 
at most one of these expectations can be zero simultaneously, and so the potential for spontaneous symmetry breaking is well-defined. 

Our new analysis does not mean that it is particularly likely that parity be spontaneously broken in QCD, since in our formulation it is 
identified as a quantum effect driven by the system dynamics through $\{V_{k}\}$. Plausibly, these symmetry breaking effects are only observed in 
instances where the statistical properties of the the Lattice ensemble become very much non quasi-classical, ie. the orders of the contributions
in (\ref{saddle}) and (\ref{branch}) become comparable \cite{hotb,hotc}.

As a second result we should notice from that it is possible to analytically continue the partition function throughout the entire $z$ plane. 
By definition the partition function is holomorphic in this region. We can deduce in principle it is therefore possible to resolve the complex
action problem via numerical simulation. If a lattice determination correctly determines the polynomial expansion in up to quantum fluctuations
at any single point in $z$, ie. that generates a covering of the Hilbert space in (\ref{Hilberta}) and (\ref{Hilbertb}), the system can be 
smoothly mapped to all point in the $z$ plane. Since the result smoothly maps to the continuum the Lattice evaluation is also sufficiently 
defined for the dynamics to generate a nonzero vacuum expectation value of the source, $\cal{O}$, and to observe spontaneous symmetry breaking 
in the continuum.

Conversely, we can understand the origin of the disconnected contributions in existing Lattice evaluations. Algorithms can be based around the 
principle that a global symmetry is exactly held, and equal Monte Carlo weights assigned for each branch. However, because only the positive 
definite contributions are kept, corresponding to real probabilistic weights, the opposite cut is in effect taken for each branch point 
multiplet in (\ref{branch1}). Genuine quantum effects can never therefore be observed as the cancellation between sectors is exact 
configuration by configuration, by construction and furthermore the system is nonrenormalizable \cite{StrongCP,4th}.

\section{Summary}
We have introduced a quantum loop operator formalism for the partition function of vector-like gauge field theories, and identified 
that this is formally equivalent to the local bosonic operator expansion on finite system size introduced in the Vafa-Witten discussion. 
The Lattice formulation is given over an extended phase space $N\otimes T$, and the source parameter is itself discretized in the new approach.
We have then considered the continuum limit of this space and identified that the zeroes of the loop operator polynomial are formally 
equivalent to partition function zeroes. Although this result is distinct from classical systems requiring a prior noncommuting limit to be taken 
over $T$ to ensure the formulation is meromorphic. The saddle point solution for the asymptotic behaviour of this complex-valued partition 
function was then determined. Spontaneous symmetry breaking is well-defined for case in which a parity symmetry is realised by the system. 
In the latter case the free energy minima can be distinct from the partition function maxima, and the generalisation of the result to general 
global symmetries was identified.

By construction, the formulation can be analytically continued across the entire phase plane in the zeroes expansion parameter. Therefore, 
any nonperturbative operator satisfying the quantum loop operator definition is an exact solution to the complex action problem.

\end{document}